\documentclass[showpacs,preprintnumbers,prb,twocolumn]{revtex4}
\usepackage{amsmath}
\usepackage{amssymb}
\usepackage{graphicx}
\usepackage{dcolumn}
\usepackage{bm}

\begin{document}

\title{The Bean model of the critical state in a magnetically shielded superconductor filament}
\author{S.V.~Yampolskii}
\email{yampolsk@tgm.tu-darmstadt.de}
\altaffiliation[On leave from ]{Donetsk Institute for Physics and Technology, National Academy of Sciences
of Ukraine, 83114 Donetsk, Ukraine}
\author{Yu.A.~Genenko}
\author{H.~Rauh}
\affiliation{Institut f\"{u}r Materialwissenschaft, Technische Universit\"{a}t Darmstadt,
D-64287 Darmstadt, Germany} 
\author{A.V.~Snezhko}
\affiliation{Materials Science Division, Argonne National Laboratory, Argonne, IL 60439,
USA}


\begin{abstract}
We study the magnetization of a cylindrical type-II superconductor filament covered by a 
coaxial soft-magnet sheath and exposed to an applied transverse magnetic field. 
Examining penetration of magnetic flux into the superconductor core of the filament on 
the basis of the Bean model of the critical state, we find that the presence of a non-hysteretic 
magnetic sheath can strongly enhance the field of full penetration of magnetic flux. 
The average magnetization of the superconductor/magnet heterostructure under consideration 
and hysteresis AC losses in the core of the filament are calculated as well.
\end{abstract}

\maketitle

\section{Introduction}

Hybrid systems composed of superconductor (SC) and soft-magnet (SM) elements 
attract much attention in view of possibilities to improve superconductor 
performance by shielding out an applied field as well as a self-induced 
transport current field~\cite{ref1a,ref1b,ref2a,ref2b}. It was first shown theoretically~\cite{ref1a,ref1b} that 
magnetic shielding may increase the critical current of a SC strip, thus 
enhancing its current-carrying capability both in the Meissner state and 
in the partly flux-filled critical state. It was found too that such shielding 
can strongly reduce the transport AC losses in SC wires and tapes~\cite{ref2a,ref2b}. 
Very intense investigations were carried out on superconducting MgB$_2$ wires 
sheathed in iron, which have become ideal objects for exploring the magnetic 
shielding effect thanks to the simplicity of their fabrication: as observed 
experimentally, such structures exhibit enhanced superconducting critical 
currents~\cite{ref3a,ref3b,ref4a,ref4b} as well as a pronounced reduction of AC losses~\cite{ref5a,ref5b} over a wide 
range of the strength of an applied magnetic field.

Recently, we have studied the flux-free Meissner state in a type-II SC filament 
covered by a cylindrical magnetic sheath and have calculated the field distribution 
as well as the magnetic moment of this composite~\cite{ref6}. Furthermore, we have considered 
the vortex state of the filament when it is exposed to a transverse magnetic field 
and/or when it carries a transport current and have derived the critical field and 
the critical current of first flux penetration~\cite{ref7,ref8} as well as the transverse lower 
critical field, $H_{c1}$ (Ref.~\onlinecite{ref8}). We found in particular that, due to the presence of 
the magnetic sheath, the field of first flux penetration can be strongly enhanced. 
Here, we concentrate on the penetration of magnetic flux into the referred composite 
when it is exposed to an applied transverse magnetic field, using the Bean model of 
the critical state~\cite{Bean}. We derive the field of full flux penetration and calculate 
both the magnetization and hysteresis AC losses in the partly flux-filled state and 
in the full critical state.

\section{Theoretical model}

Let us consider an infinitely extended, cylindrical type-II SC filament of radius $R_1$ 
covered by a coaxial magnetic sheath of thickness $d$ and relative magnetic permeability $\mu \gg 1$; 
the structure is aligned parallel to the $z$-axis of a Cartesian coordinate system $x,y,z$ 
adapted to the filament (see Fig.~1). 
\begin{figure}[!bp]
\includegraphics[height=7.5cm]{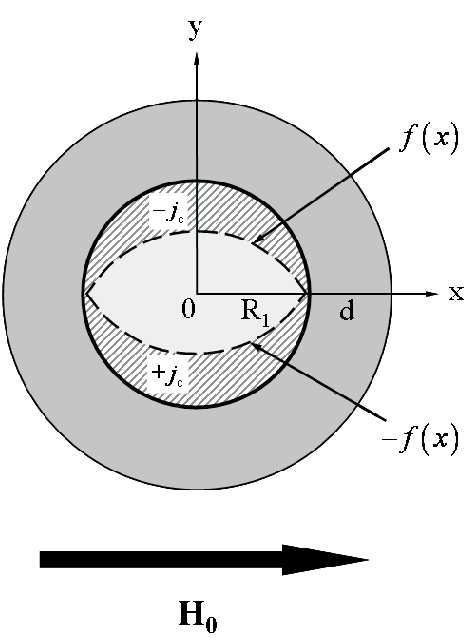}
\caption{Cross-sectional view of the SC filament of radius $R_1$ (light region) 
and the coaxial magnetic sheath of thickness $d$ (dark region), their axes 
coinciding with the $z$-axis of a Cartesian coordinate system $x,y,z$. The 
solid ring indicates an insulating layer between the SC core and the 
magnetic sheath; the dashed curves indicate the boundaries between the 
flux-penetrated regions (shaded areas) and the region of the Meissner state. 
The direction of the applied magnetic field $H_0$ is also marked.}
\end{figure}
A transverse magnetic field $H_0$ is applied along 
the positive $x$-direction and is asymptotically uniform at distances large compared to $R_2=R_1+d$. 
Magnetic flux enters the SC core through the SC/SM interface and induces shielding currents 
in the regions of flux penetration. According to the Bean concept of the critical state, 
the flux-penetrated regions carry a current of density $j_c$ directed parallel (antiparallel) 
to the $z$-direction, as marked in Fig.~1. The distribution of the magnetic field in the composite 
is governed by the Maxwell equations
\begin{equation}
\bf {\nabla} \times {\bf H} = 0,  \qquad  	 \bf {\nabla} \cdot {\bf B} = 0.
\label{Maxwell}	
\end{equation}
Invoking the existence of an insulating nonmagnetic layer between the superconductor and 
the magnetic sheath, of thickness much less than $d$, $R_1$ and the London penetration depth $\lambda$ 
(as observed, e.g., in MgB$_2$/Fe composites~\cite{ref4a,ref4b}), we supplement equations~(1) with boundary 
conditions implying continuity of the tangential component of the magnetic field as well as 
of the normal component of the magnetic induction when the interface between the superconductor 
and the magnet as well as the surface of the magnet adjacent to vacuum are traversed. 
In addition, we require the magnetic field to approach asymptotically the value of the 
applied magnetic field, $H_0$.

In our consideration, we take no account of the effects of remanent magnetization, 
non-linear dependence of the magnetization on the magnetic field and finite conductivity 
of the magnetic layer, regarding the (scalar) relative permeability $\mu$ as the only characteristic 
of the sheath. With these assumptions, we represent the dependence of the magnetic induction ${\bf B}_M$ 
on the magnetic field ${\bf H}_M$ in the sheath as split in two regions separated by the value of 
the saturation field, $H_s$: if $ H_M \le H_s$, we approximate ${\bf B}_M$(${\bf H}_M$) by a linear dependence,
${\bf B}_M^{<} = \mu_0 \mu {\bf H}_M$; 
conversely, if $ H_M \ge H_s$, the value of the magnetization saturates, and the magnetic induction 
takes the form ${\bf B}_M^{>} = \mu_0 \left[ {\bf H}_M + \left( \mu -1 \right) {\bf H}_s \right]$. 
We then also assume that ${\bf B}_M$, ${\bf H}_M$ and ${\bf H}_s$ are collinear.

In the geometry addressed, the magnetostatic problem is to determine a contour curve $y = f \left( x \right)$  
which bounds the region occupied by the critical state so that inside the contour the magnetic 
induction ${\bf B}_{SC} = 0$, i.e. the inner region of the SC is kept in the flux-free state. 
The magnetic induction depends on the $x$- and $y$-coordinates everywhere, and in the region 
of vanishing current its components satisfy the Laplace equation $\nabla^2 B_{\alpha,SC} =0$ ($\alpha = x,y$). 
By virtue of the extremum theorem of complex analysis~\cite{ref10}, it is sufficient, 
then, to determine $f \left( x \right)$   
such that the condition ${\bf B}_{SC} = 0$ is satisfied identically along the curve $y = f \left( x \right)$.

\section{Results and discussion}

\subsection{Partly flux-filled critical state of the SC core of the filament}

We first look at the situation when magnetic flux only partly penetrates the SC core 
of the filament, the situation depicted in Fig.~1. To facilitate further considerations, 
we refer to an assumption invoked in previous works~\cite{ref5a,ref5b}, viz. that the total magnetic 
induction in the SC core can be decomposed according to
\begin{equation}
{\bf B}_{SC} = {\bf B}_{i} + {\bf B}_{e}, 
\label{eq2}
\end{equation}
\noindent where ${\bf B}_{i}$ is the induction due to the current distribution in the pure SC filament and 
${\bf B}_{e}$ is the induction in the hole of the magnetic sheath in the absence of the SC core, 
neglecting the effect of the SC in the magnetic sheath. The task, therefore, is to 
determine the contour curve $y = f \left( x \right)$ such that the current distribution generates a {\it uniform} 
magnetic induction ${\bf B}_{i}$ in the region enclosed by the contour which cancels the induction 
${\bf B}_{e}$ so as to yield ${\bf B}_{SC} = {\bf B}_{i} + {\bf B}_{e} = 0$.

The induction ${\bf B}_{i}$ can be determined numerically following a previously suggested approach~\cite{Ashkin}. 
Figure~2 shows the so-calculated contour delineating the flux-penetrated region for different 
values of the normalized magnetic field $ h_e = 4 \pi B_e / \mu_0 j_c R_1$. 
\begin{figure}[!tbp]
\includegraphics[width=8cm]{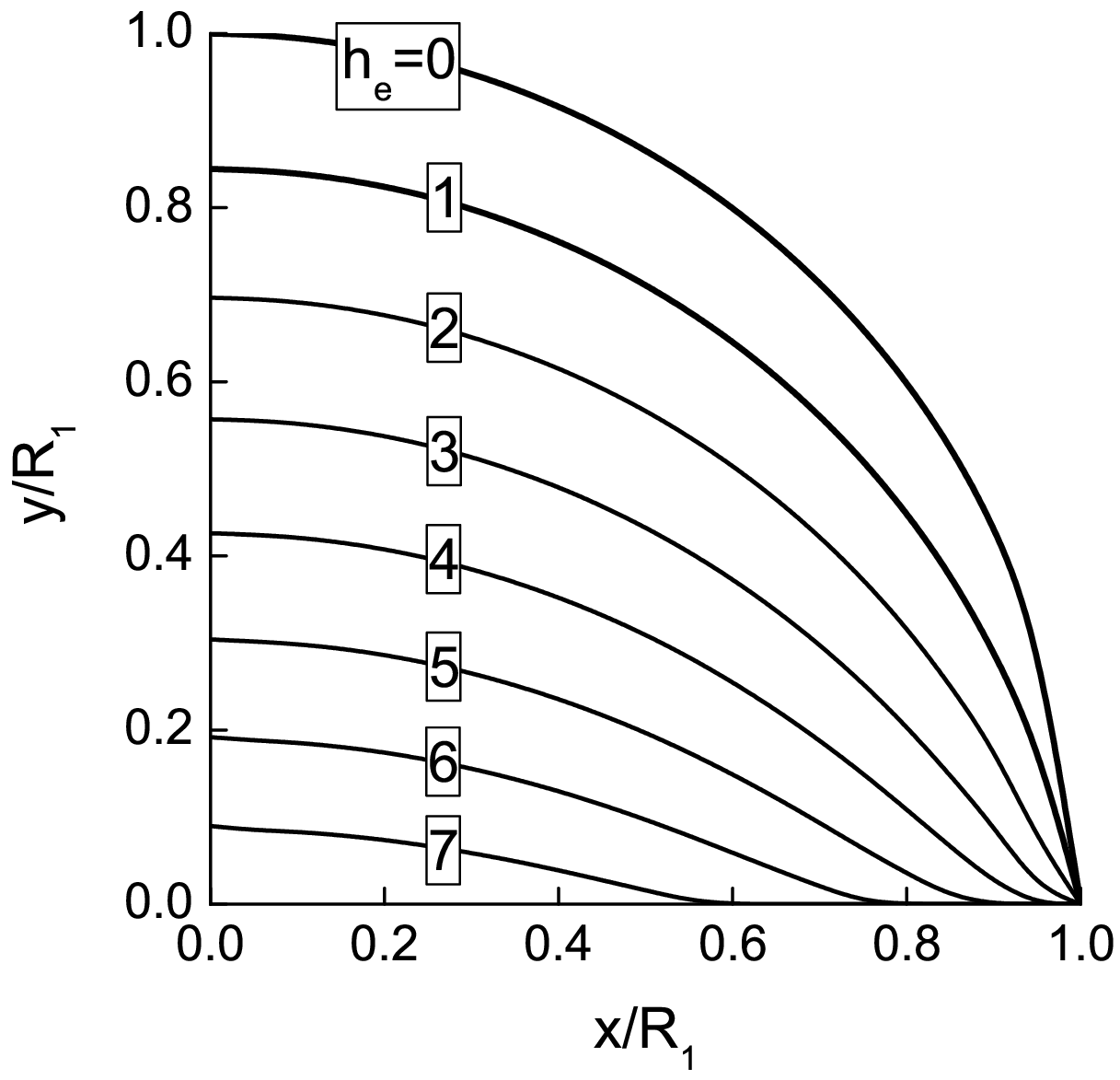}
\caption{Contours separating flux-penetrated regions (first quadrant) for a range 
of values of the normalized magnetic field $h_e = 0(1)7$.}
\end{figure}
Due to a finding from numerical analysis, full penetration of magnetic flux into the SC core 
takes place for $h_e = 8$; a value which is in perfect agreement with the well-known result 
$B_p^0 = 2 \mu_0 j_c R_1/ \pi$ for 
the {\it pure} SC filament~\cite{Ashkin,ref12,ref13}. Figure~3 depicts the calculated dependence of the diamagnetic 
moment of the SC core, per unit length of the filament, $M_{SC}$, on the magnetic induction $B_e$. 
Obviously, $M_{SC}$ reaches its extremum value when the magnetic flux completely penentrates the SC core.
\begin{figure}[!tbp]
\includegraphics[width=8cm]{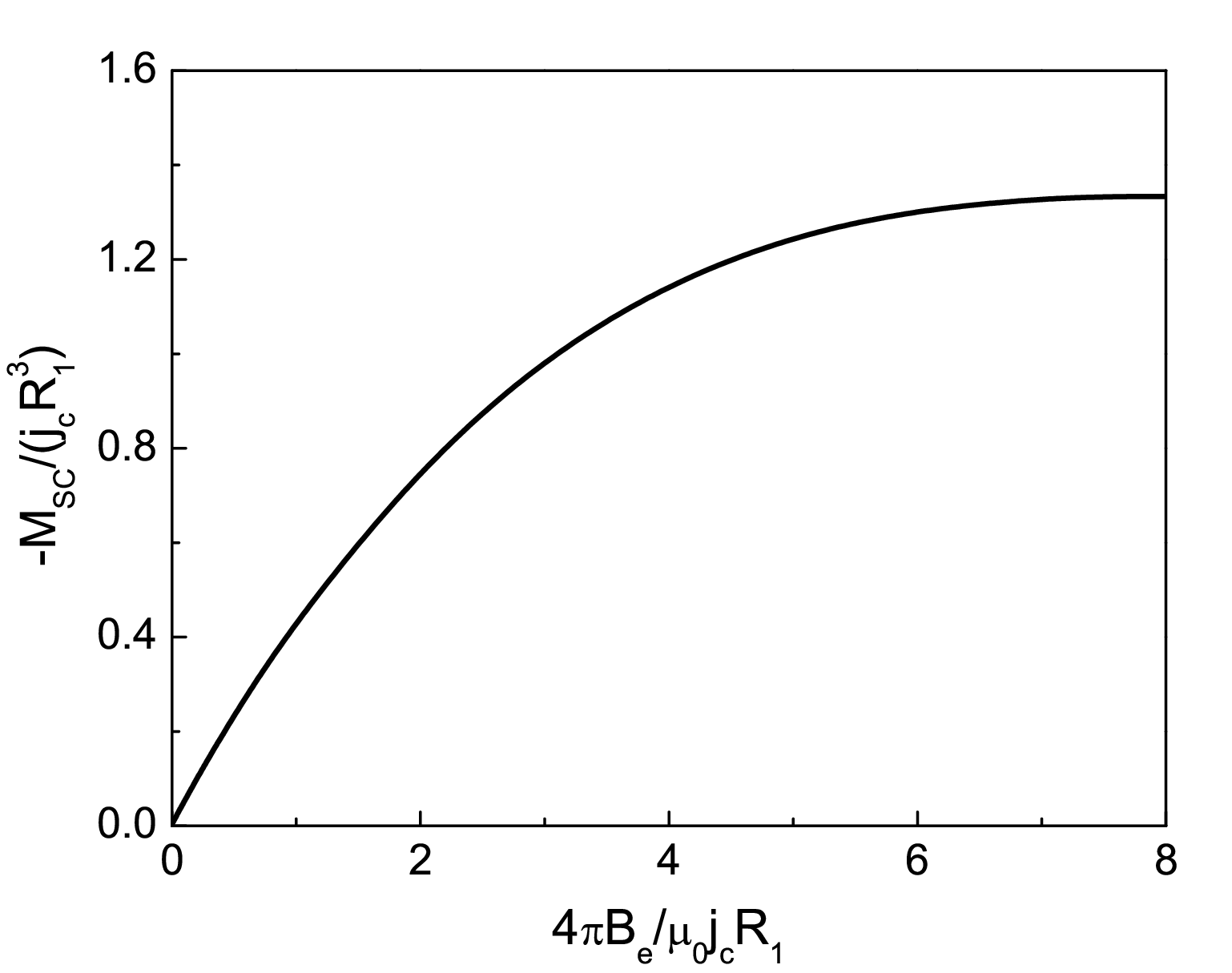}
\caption{Dependence of the diamagnetic moment of the SC core, per unit length of the filament, 
$M_{SC}$, on the magnetic induction $B_e$.}
\end{figure}

As is well known, the dependence of the magnetic induction $B_e$ on the applied magnetic field, 
if $ H_M \le H_s$, equals $B_e = \alpha^< \mu_0 H_0$ with
\begin{equation}
\alpha^< = \frac{4 \mu}{\left( \mu +1 \right)^2 - \left( \mu -1 \right)^2 R_1^2/R_2^2}	
\label{eq3}
\end{equation}
(see, e.g., Ref.~\onlinecite{ref14}). If $ H_M \ge H_s$, the magnetic induction $B_e$ can be represented in the form 
$B_e = \mu_0 \left( H_0 - \alpha^> H_s \right)$, where
\begin{equation}
\alpha^> = \frac{\left( \mu -1 \right)^2 \left( 1- R_1^2/R_2^2 \right)}{\left( \mu +1 \right)^2 - \left( \mu -1 \right)^2 R_1^2/R_2^2}.	
\label{eq4}
\end{equation}
From these results we obtain the field of full flux penetration,
\begin{align}
H_p^< & = B_p^0/ \alpha^< \mu_0, \label{eq5} \\
H_p^> & = B_p^0/\mu_0 + \alpha^> H_s,	
\label{eq6}
\end{align}
\noindent for the non-saturated and the saturated state of the magnet, respectively. Apparently, 
the field $H_p$ is most strongly influenced by the magnetic sheath when the magnet is in the 
non-saturated state, growing linearly with $\mu$ in the limit $\mu \to \infty $. When the magnet is in the saturated state, 
the dependence of the field $H_p$ on the parameters of the magnet is less pronounced, reaching the maximum value   
$B_p^0/ \mu_0 + H_s $ in the limit $\mu \to \infty $.

\subsection{Full critical state of the SC core of the filament}

When the magnetic flux completely penetrates the SC core, the boundary between the regions 
permeated by currents of opposite signs coincides with the line $y = 0$. For this situation, 
the field of full flux penetration as well as the magnetic moment of the composite can be 
calculated exactly without invoking assumption~(2). The field $H_p$ is
\begin{align}
H_p^< & = \frac{B_p^0}{\mu_0} \left[1+ \frac{\left( \mu -1 \right) \left( 2\mu -1 \right) 
\left( 1-R_1^2/R_2^2 \right)}{6 \mu} \right], \label{eq7} \\
H_p^> & = \frac{B_p^0}{\mu_0} \left[1 + \alpha^> \frac{\mu +1}{3 \left( \mu +1 \right)} \right]
+ \alpha^> H_s,	\label{eq8}
\end{align}
\noindent for the non-saturated and the saturated state of the magnet, respectively. 
The magnetic moment of the magnetic sheath for the non-saturated state of the magnet, 
per unit length of the filament, which is determined by the surface integral of the field 
$ {\bf H}_M$ taken over the cross-section of the magnetic sheath, reads
\begin{equation}
M^<_M = \frac{2 \left( \mu^2 -1 \right) H_0 + \left( \mu -1 \right)^2 \left( 2 B_p^0 R_1^2/3 \mu_0 R_2^2 \right)}
{\left( \mu +1 \right)^2 - \left( \mu -1 \right)^2 R_1^2/R_2^2}.
\label{eq9}
\end{equation}
The value of this moment saturates at the field $H_0 = H_s$, i.e. it does not change at 
higher applied magnetic fields. The diamagnetic moment of the SC core, per unit length 
of the filament, is $M_{SC}= -4 j_c R_1^3/3$; a value which coincides with the extremum value of the calculated 
diamagnetic moment displayed in Fig.~3.

A comparison of results shows that the field $H_p$ for the full critical state, equations (7) and (8), 
turns out to be always larger than the respective field for the partly flux-filled critical state, 
equations (5) and (6), the difference amounting to about 25\% in the limit $\mu \to \infty $. 
The SC core of the filament hence adopts the full critical state for an applied magnetic field 
with a strength larger than that following from considerations using the assumption~(2). 
This shows that the latter assumption only roughly models the real situation, since it ignores 
the effect of the increase of the magnetic flux density in the magnetic sheath due to flux 
expulsion from the SC core; a fact first established for the virgin magnetization of a SC/SM filament~\cite{ref6}.

\subsection{Hysteresis AC losses}

When the SC/SM composite is magnetized by an alternating magnetic field with maximum amplitude $H_m$, 
hysteresis losses are bound to occur. The numerically calculated magnitude of the loss of energy, $U$, 
can be fitted to the empirical expression~\cite{ref12}
\begin{equation}
U = \frac{2 \mu_0 H_p^2}{3} \left\{
\begin{aligned}
& \left( \frac{H_m}{H_p} \right)^3 \left( 1 - \frac{H_m}{2H_p} \right), & \quad H_m < H_p, \\
& \frac{H_m}{H_p} -\frac{1}{2}, & \quad H_m \ge H_p.\\
\end{aligned} 
\right.
\end{equation}
This reveals that the dependence of hysteresis AC losses on the parameters of the magnetic sheath – 
its thickness and relative permeability – is completely determined by the field of full flux 
penetration, $H_p$.

\section{Summary}

In conclusion, we have studied the magnetization of a cylindrical type-II SC filament covered by a coaxial magnetic sheath and exposed to an applied transverse magnetic field. We have calculated the field of full flux penetration into the SC core, the magnetic moment of the composite in the full critical state as well as the diamagnetic moment of the SC core in the partly flux-filled critical state. We have shown that the field of full flux penetration increases due to the shielding of the applied field by the magnetic sheath and due to the expulsion of magnetic flux from the SC core. We argue that the latter phenomenon, which has been ignored in previous considerations, must be taken into account to correctly describe the magnetic response of such a composite, especially when it is in the partly flux-filled critical state; a problem which will be addressed in future considerations.

\begin{acknowledgments}
This work was supported by a research grant from the German Research Foundation (DFG).
\end{acknowledgments}

\bibliographystyle{plain}
\bibliography{apssamp}

\end{document}